\documentstyle[12pt,a41,epsfig,wrapfig]{article}

\begin{document}
\begin{flushleft}
DESY 98-139\\
hep-ph/9809413\\
September 1998
\end{flushleft}
\vspace*{4cm}

\begin{center}
{\Large \bf Double Spin Asymmetries in $P$-wave Charmonium  Hadroproduction}

\vspace*{2cm}
{W.-D.~NOWAK, A. TKABLADZE\footnote{Alexander von Humboldt Fellow}}\\
\vspace*{1.cm}
{\it DESY Zeuthen, D-15738 Zeuthen, Germany}\\

\vspace*{2.5cm}

\end{center}

\begin{abstract}
\noindent
{\small
We discuss the   double spin asymmetries in $P$-wave charmonium hadroproduction
 with non-zero transverse momenta at
fixed target energies, $\sqrt{s}\simeq40$ GeV, within the framework of the
factorization approach.
The size of the  asymmetries and the projected statistical errors
in a future option of HERA with longitudinally
polarized protons scattering off a polarized target (HERA-$\vec N$) are 
calculated. Measurements of the $\chi_{c1}$ and $\chi_{c2}$ decays into 
dilepton plus photon
 should allow  to
distinguish between different parametrizations for the polarized gluon
distribution in the proton.
At higher energies ($\sqrt{s}=200$ GeV) the situation appears less favourable with the presently envisaged integrated luminosities of the polarized RHIC 
collider. 
}
\end{abstract}

\newpage
\section{Introduction}
\setcounter{equation}{0}
\vspace{-3mm}

The study of spin asymmetries in the production of 
heavy quarkonium states  in
polarized nucleon-nucleon collisions should provide  important
information about the  spin structure of the nucleon. Heavy quark-antiquark
production processes occur at  small distances and the subprocess level
cross sections can be calculated perturbatively.    
On one hand, charmonium production asymmetries are expected to be
sensitive to the polarized gluon distribution function in the proton,
since  heavy quark systems  are mainly produced in gluon-gluon fusion
subprocesses.
On the other hand, it is essential  to  investigate in more
detail  the heavy quark-antiquark pair hadronization phase. To this
end,  observation of charmonium production in experiments with polarized
beams is expected to  provide  additional tests for existing models.

The Factorization Approach (FA) based on the Nonrelativistic QCD (NRQCD) 
turns out to be a most rigorous framework to study heavy quarkonium production
and deacy processes \cite{NRQCD}. According to the FA the
 inclusive production
cross section for a quarkonium state $H$ in the process 
\begin{eqnarray}
 A+B\to H+X
\end{eqnarray}
can be factorized as
\begin{eqnarray}
\sigma(A+B\to H) = 
\sum_{n}{\frac{F_n}{m_Q^{d_n-4}}\langle0|{\cal O}^H_n|0\rangle},
\end{eqnarray}
where the short-distance coefficients, $F_n$, are associated with the
production of the  heavy quark pair in the color and angular momentum state 
$[n]$. This part of the 
production cross section involves only momenta of the order of the 
heavy quark
mass and larger and can be calculated perturbatively. The heavy quark-antiquark
 pair production process occurs at  small distances, $~(1/m_Q)$, and
is  factorized from the hadronization phase which takes place at large 
distances, $1/(m_Q v^2)$, where $m_Q$ is the heavy quark mass.
 Here $v$ is the average velocity of heavy
 constituents in the 
quarkonium, with  $v^2\simeq0.3$ for charmonium and $v^2\simeq0.1$ for 
bottomonium systems. 
The vacuum matrix elements of NRQCD operators, $\langle0|{\cal O}^H_n|0\rangle$
\cite{NRQCD}, describe the evolution of the quark-antiquark state $[n]$ 
into the final hadronic state $H$.
 These matrix elements cannot
be calculated perturbatively, but the relative importance
 of long distance matrix elements in powers of velocity $v$ can be
 estimated  using
the NRQCD velocity scaling rules \cite{LMNMH}.
This   formalism implies that quark-antiquark color octet intermediate
 states are allowed
to contribute to heavy quarkonium  production and decay
processes at higher order in the velocity expansion.
Therefore, in the FA  the complete structure of
the quarkonium Fock space is taken into account while in the old approach,
in the  Color Singlet Model (CSM) \cite{CSM},
only the dominant Fock state is considered, which consists of a color singlet
quark-antiquark pair in a definite angular-momentum state of the final
hadron (the leading term in the velocity expansion).

The predicted 
shape of the  $p_T$ distribution of the $^3S_1^{(8)}$ intermediate 
octet state production cross section indicates that $J/\psi$ and $\psi'$
production at large $p_T$ observed at the Tevatron (FNAL)
can be explained in the FA \cite{BF,CL}.  
Recent investigations have shown  that  the contribution of color octet states 
to the charmonium production cross section is very important at fixed 
target energies, $\sqrt{s}\simeq30-60$ GeV,
 and reduces existing discrepancies between experimental data and 
predictions of the CSM \cite{FixTag}.

Despite the obvious successes of the NRQCD in explaining large $p_T$
charmonium production at the Tevatron, some experimental data contradict 
 the Color Octet Model (COM) predictions.
 In particular, theoretical
predictions disagree  with measurements of the polarization of
   $J/\psi$ and $\psi'$ particles produced at
fixed target energies \cite{BeR} and  the COM prediction for the
yield ratio of $\chi_{c1}$ and $\chi_{c2}$ states remains too low \cite{BeR}.
One  possible solution for  these discrepancies was proposed
by Brodsky et al. \cite{VHBT}  suggesting that higher twist 
processes, when more than one parton from projectile or target participate
in the reaction, might give a significant 
contribution to low $p_T$ production of $J/\psi$ and $\chi_{c1}$ states.
Problems exist also in charmonium photoproduction at HERA.
The color octet contribution underestimates the inelastic $J/\psi$ 
photoproduction cross section at large values of $z$
($z=E_{J/\psi}/E_{\gamma}$ in the laboratory frame) \cite{CK}.

 Unlike the color singlet long distance matrix elements, each connected
 with the subsequent
hadronic non-relativistic wave function at the origin, color octet long
distance matrix elements are unknown and have to be extracted from
experimental data.
The NRQCD factorization approach implies universality, i.e. the values
of long distance matrix elements extracted from  different
experimental data sets must be the same. 
However, due to the presently rather large theoretical 
uncertainties \cite{Trig,BK} and the unknown size of  higher twist 
process contributions \cite{VHBT} the existing  experimental data does
 not allow  to check the FA universality, yet.

This fact motivated us to look for other processes with less
theoretical uncertainties to test the COM.
The observation of spin asymmetries in the production of charmonium states 
can be used for these purposes \cite{TT}
as well as  measurements of  the $J/\psi$ polarization
in  unpolarized hadron-hadron collisions and electroproduction 
\cite{VHBT,BK,FleN}.

In the present article we consider double spin asymmetries in $P$-wave 
charmonium production at non-zero $p_T$ in the NRQCD factorization approach.
The double spin asymmetry in $J/\psi$ hadroproduction has been studied 
recently taking into account color octet intermediate states \cite{TT,GM}.
The asymmetries in the production of $\chi_{cJ}$ states
at small $p_T$ were considered in refs. \cite{GM,Don}. 
As already  mentioned above, at small values of $p_T$  large contributions 
from  higher twist effects are expected \cite{VHBT} and the presently 
available theoretical predictions are not reliable enough 
 to allow extraction of information about polarized parton distributions 
or to check the COM.
 Morii et al.  considered asymmetries in the production 
of $\chi_{cJ}$ states  at RHIC energies taking into account
 only  fragmentation 
type contributions for color octet intermediate states \cite{MTY}.
However, since the cross section for 
  the production of charmonium states falls steeply  
 with increasing $p_T$  the  statistical errors
 becomes too large at those  values of $p_T$ where the  fragmentation 
approach can be applied, especially at at small energies 
$\sqrt{s}\simeq50$ GeV.
In contrast to this approach, our calculations \cite{NTT} use
  the exact expressions
for the cross sections of   intermediate color octet states  
in different helicity states of the
initial partons,  to estimate the expected spin asymmetries 
in the production of 
 $\chi_{cJ}$ states  as 
well as the projected errors.
 
The article is organized as  follows. In the next section  
subprocess level asymmetries for possible intermediate octet and singlet 
states and corresponding long-distance color octet parameters are considered.
In section 3  numerical results are presented for spin asymmetries in 
production of $\chi_{cJ}$ states  at HERA-$\vec N$ \cite{Nowak}, one of the 
future options of HERA;
an experiment using an internal polarized nucleon target in  a possibly later 
polarized HERA beam at  $\sqrt{s}\simeq40$ GeV. 
For comparison, we also consider the expected  asymmetries  in the production
of $\chi_{cJ}$ states  in  similar spin physics experiments
at the RHIC collider \cite{RHIC} at $\sqrt{s}=200$ GeV.

\vspace{-3mm}

\section{Asymmetries at the Subprocess Level}

The two-spin asymmetry $A_{LL}$ for  inclusive $\chi_{cJ}$ 
  production  is defined as
\begin{eqnarray}
A_{LL}(pp) = \frac{
d\sigma(p_+p_+\to \chi_{cJ})-d\sigma(p_+p_-\to \chi_{cJ})}
{d\sigma(p_+p_+\to \chi_{cJ})+d\sigma(p_+p_-\to\chi_{cJ})}=
\frac{Ed\Delta\sigma/d^3p}{Ed\sigma/d^3p}.
\end{eqnarray}
where the subscript in $p_+(p_-)$ indicates the sign of the 
helicity projection onto the direction of the proton.
The production of each quarkonium state receives  contributions from both  
color octet and color singlet states.
In  leading order  $v^2$ only one color octet state contributes to the
 production of $\chi_{cJ}$ states, namely $^3S_1^{(8)}$.
Unlike direct $J/\psi$ production, leading color octet and color singlet
 contributions to P-wave charmonium production scale equally in $v^2$, 
$O(v^5)$, and the subleading corrections are only of the order $O(v^9)$.
To calculate the production asymmetries  only leading order
 color octet contribution are taken into account.
\begin{figure}[ht]
\vspace{-4mm}
\centering
\epsfig{file=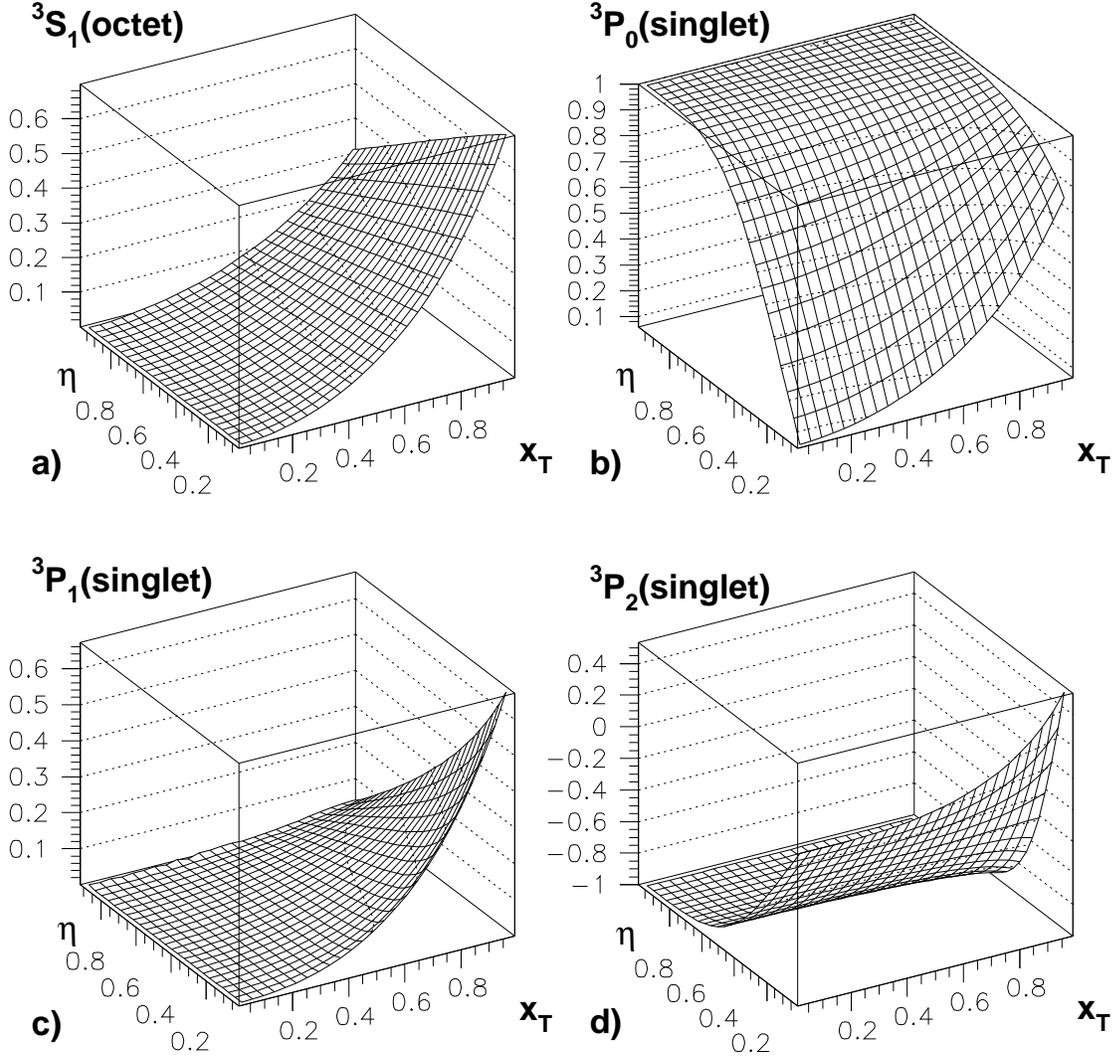, width=16.5cm}
\caption{\small
  The subprocess level asymmetries for $^3P_0$, $^3P_1$, and $^3P_2$ color
 singlet, as well as for $^3S_1^{(8)}$ color octet state.}
\end{figure}

We consider the production of $\chi$ states  at non-zero 
values of transverse momentum, $p_T>1.5$ GeV.
The leading contribution to the production of charmonium states
at those  values of 
transverse momentum comes from the  $2\to2$ subprocesses.
The leading order
 subprocesses, $2\to1$, should not contribute to
 the quarkonium production with $p_T>1.5$ GeV because 
these  transverse momenta cannot be caused by internal motion of
partons in the nucleon. The contribution of  higher twist processes
\cite{VHBT}  is also expected to be small at $p_T>1.5$ GeV.

For the calculation of the expected asymmetries for color octet and singlet
states of a $(c\bar c)$-pair we consider  the following
subprocesses:
\begin{eqnarray}
 g+g &\to& (c\bar c)+ g\nonumber \\
 g+q &\to& (c\bar c)+ q           \\
 q+\bar q&\to& (c\bar c)+g\nonumber
\end{eqnarray}
The asymmetries for all these subprocesses were calculated in ref. \cite{TT}.
The values of those giving a significant contribution to the production of 
$\chi_{cJ}$ states are presented in Fig. 1 in dependence 
on  the two dimensionless
quantities $\eta=4 m_c^2/\hat s$ and $x_T=p_T/p_{max}$, where
$m_c$ is the charm quark mass, $\hat s$ the c.m. energy in 
the parton-parton system, and
 $p_{max}$
is the maximum momentum of the produced state in the subprocess.
  Figure 1    shows only the gluon-gluon fusion subprocess asymmetries
because they give the main contribution to the hadron level asymmetry.

The color singlet long distance matrix elements are connected to quarkonia 
wave functions at the origin:
\begin{eqnarray}
\langle0|{\cal O}_1^{\chi_{cJ}}(^3P_J)|0\rangle = 
\frac{3 N_c}{2\pi} (2 J+1)|R'(0)|^2=3.2\cdot10^{-1} GeV^3.
\end{eqnarray}
This  value   was used in \cite{CL} for 
fitting the CDF data on   $J/\psi$ production through $\chi_{cJ}$ states 
and corresponds to the Buchm\"uller-Tye type potential solution tabulated in
 \cite{EQ}. 
For the color octet long distance matrix element the following value
extracted from CDF data was used \cite{CL}:
\begin{eqnarray}
\langle0|{\cal O}_8^{\chi_{c1}}(^3S_1)|0\rangle &= &9.8\cdot10^{-3} GeV^3,
\end{eqnarray}
\begin{wrapfigure}{l}{7.5cm}
\vspace*{-10mm}
\centering
\epsfig{file=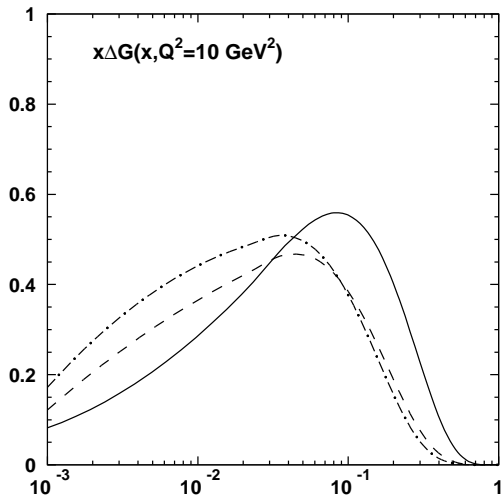,width=7.5cm}
\vspace{-7mm}
\caption{\small 
Different possible polarized gluon
 distributions in the nucleon, used in this paper:
 GS parametrization NLO set A (solid line),
NLO set B (dashed line)   and  LO set A (dash-dotted line) \cite{GSnew}.
}
\end{wrapfigure}
As shown in ref. \cite{BK} the variation
of the renormalization and/or the factorization should  lead to 
large uncertainties when  fitting the color octet parameters. 
The fit results for  long distance color octet matrix elements
can be affected also by higher $v^2$ corrections (so called 
`trigger bias' effect) \cite{Trig}.
Therefore  care is required when using  at fixed target energies the 
value for this parameter
 extracted  from  CDF.
The influence of these uncertainties on the expected 
asymmetries will be discussed in the next section.

\vspace{-3mm}

\section{Results and Discussion}
\vspace{-3mm}

 The characteristic value of the partonic $x$ in the production of
$(c\bar c)$ pairs  can  be obtained from the relation
$x_1 x_2\simeq(4 m_c^2+p_T^2)/s$ ($\simeq 0.01$ at HERA-$\vec N$). 
This means that the typical value of $x_{gluon}$ that
can be probed by
 measuring the spin
 asymmetry in charmonium
  production is about  $x_{gluon}\simeq0.1$. 
We used three  different sets of Gehrmann and Stirling (GS) 
parametrizations \cite{GSnew} for polarized parton distribution 
functions (PDF) that are different in the region $x\simeq0.1$
to show the sensitivity  of the spin asymmetries in the
 production of  $\chi_{cJ}$
states on the gluon polarization in the nucleon: the sets A and B of 
the NLO GS parameterization  and the LO parameterization set A.
 In  Fig. 2 the polarized gluon densities from these
parametrizations are shown  at $Q^2=10$ GeV$^2$, which is the
 appropriate value in the  kinematical region considered.
As can be seen, the three chosen sets exhibit somewhat different 
values for the polarized gluon distribution function
at partonic $x$ values around $0.1$.
We note that, although the calculations of subprocess
level cross sections are performed in leading order, 
the  NLO set of the parametrization was
used    to probe  different shapes of the polarized gluon distribution.
\begin{figure}[ht]
\vspace{-4mm}
\centering
\begin{minipage}[c]{7.5cm}
\centering
\epsfig{file=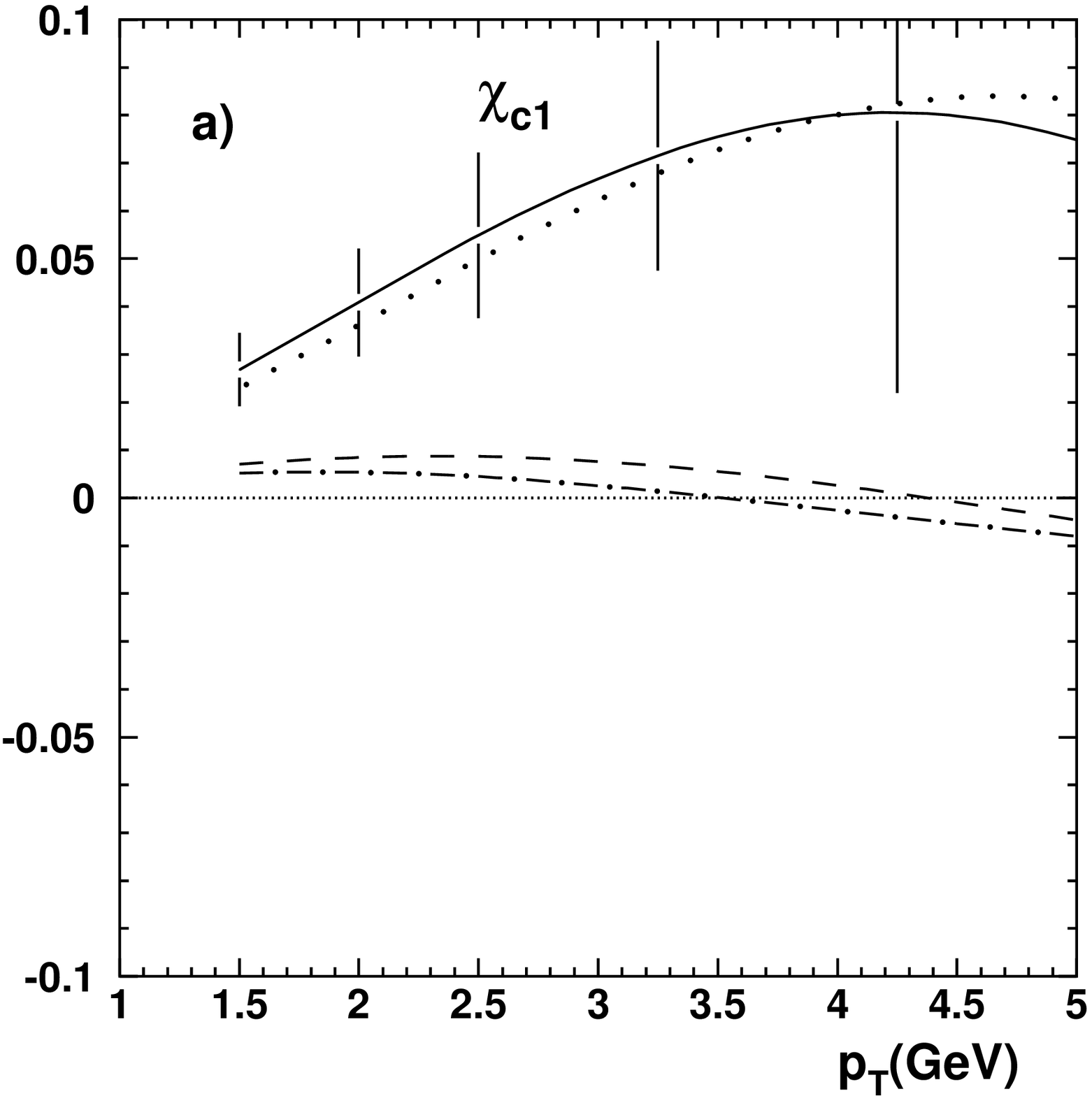,width=7.5cm}
\end{minipage}
\hspace*{0.5cm}
\begin{minipage}[c]{7.5cm}
\centering
\epsfig{file=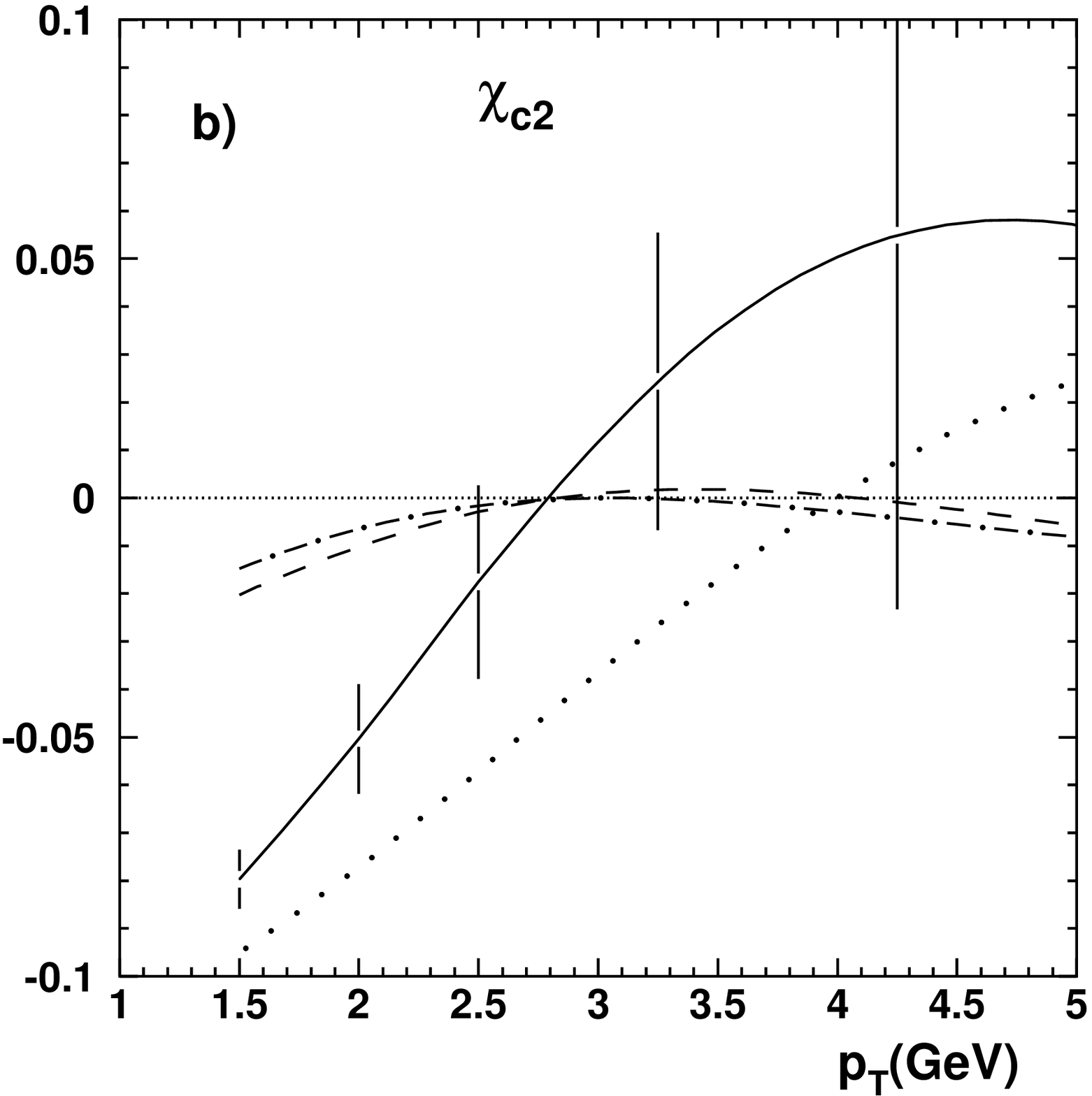,width=7.5cm}
\end{minipage}
\caption{\small
  a) $\chi_{c1}$ and  b) $\chi_{c2}$ production  asymmetries
 versus transverse momentum 
at  $\sqrt{s}=40$ GeV.
 The solid (dashed) line correspond
to set A (B) of the NLO GS parametrization
 and the dash-dotted line to set A of the 
leading order GS parametrization.
The dotted line represents the expected asymmetry calculated 
only in the CSM for the set A of the NLO parametrization.
}
\end{figure}

Figs. 3 and 4  show the expected production 
asymmetries for  $\chi_{cJ}$ states at $\sqrt{s}=40$ GeV (HERA-$\vec N$).
Figures  3a and 3b also include  projected statistical uncertainties 
on the asymmetry $\delta A_{LL}$, estimated  at HERA-$\vec N$
 from \cite{NowNew} by
\begin{eqnarray}
\delta A_{LL} = 0.17/\sqrt{\sigma(pb)};
\end{eqnarray}
where $100\%$ efficiency is assumed.
This relation is based upon an integrated luminosity of
$240~pb^{-1}$ and beam and target polarizations of $P_B=0.6$ and 
$P_T=0.8$, respectively \cite{Nowak}.
The error bars  are obtained by using integrated cross sections over bins
$\Delta p_T=0.5$ GeV (for the first three points) and $\Delta p_T=1$ GeV (for
the other two ones).
Note that it is easier to measure asymmetries for $\chi_{c1}$ and 
$\chi_{c2}$ states due to their  large branching ratios for decays
 into 
$J/\psi$  plus photon ($27\%$ for $\chi_{1c}$ and $13.5\%$ for
 $\chi_{2c}$).
These  branching ratios  and the
$J/\psi$ decay branching ratios into  $l^+l^-$ ($l=e,\mu$) 
are taken into account in  the
calculations of the projected statistical  errors for the  production
of $\chi_{c1}$ and $\chi_{c2}$
states.
Due to the small branching ratio of the $\chi_{c0}$  decay through $J/\psi$,
$6.6\cdot10^{-3}$, the projected statistical errors of an asymmetry measurement
for this state are too large.

In Figs. 3 and 4, the  asymmetries for set A and set B of the NLO GS parametrization are
 represented by solid and dashed lines, respectively, whereas 
the dash-dotted lines correspond to set A of the LO parametrization.
 The  asymmetries coming only from color 
singlet intermediate states  expected for the set A of  NLO GS
parametrization are additionally shown in Figs. 3 and 4 by dotted lines.
Note that for the set C of the NLO GS
parametrization all asymmetries are less than $1\%$ and 
practically unobservable, hence they are not shown.
For the mass of the charm quark the value $m_c=1.48$ GeV was taken and
the parton
distribution functions are evaluated on the factorization  scale
$\mu = \sqrt{p_T^2+4 m_c^2}$.  The strong coupling constant is calculated
by the one-loop formula with 4 active flavors ($\Lambda_{QCD}=200$ MeV).

In all three parameterizations for the polarized parton distribution
functions the
gluon-gluon fusion process gives the dominant contribution 
to  $\Delta\sigma$ in the 
 kinematical region considered. The 
quark-gluon subprocesses contribute only to about $10\%$ 
(for  both color octet and color singlet 
states) and the contribution of quark-antiquark annihilation subprocesses 
is less than $1\%$. 

\begin{wrapfigure}{l}{7.5cm}
\centering
\epsfig{file=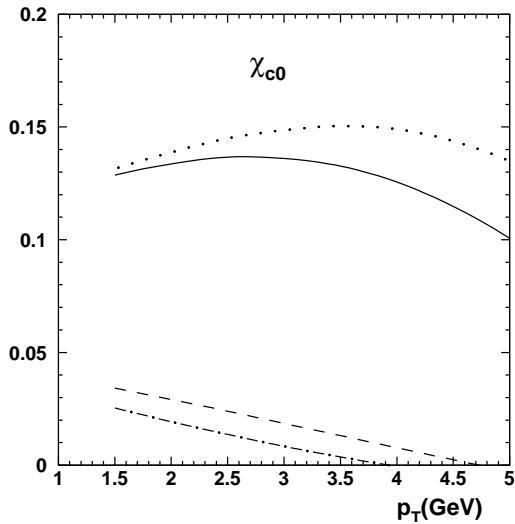,width=7.5cm}
\caption{\small
Expected asymmetries for $\chi_{c0}$ production at HERA-$\vec N$
for different GS parameterizations as explained in Fig. 3.
}
\end{wrapfigure}

The contribution of the color octet $^3S_1^{(8)}$ state does practically 
 not change  the $\chi_{c1}$ production asymmetry coming from the color singlet
 intermediate state. For $\chi_{c0}$, the influence of the 
 octet contribution becomes larger
 with increasing $p_T$.
 As seen from  Fig.~1, the asymmetries for 
color singlet $^3P_2$ and color octet $^3S_1^{(8)}$ states
have different sign.
By this reason the color octet contribution,
compared to the color singlet one,
 changes the asymmetry of the $\chi_{c2}$ state 
 more than in the case
 of $\chi_{c0}$ and $\chi_{c1}$.
Moreover, due to the `trigger bias' effect the corresponding color octet long 
distance matrix element may be approximately two  times larger and hence
the contribution of the color octet states to the asymmetries may even
increase.
 
In the ~inclusive ~case ~the ~kinematics of ~the ~$2\to2$ subprocess cannot
be reconstructed completely. Hence only indirect
information on the  gluon polarization of the nucleon
can be obtained by measuring the spin asymmetry in 
 inclusive $\chi_{cJ}$ production. As can be seen from Figs. 3 and 4, the 
expected asymmetries for all $\chi_c$ states strongly depend on the size 
of the polarized gluon distribution function in the region  $x_{gluon}\ge0.1$.
At HERA-$\vec N$ the size  of the $\chi_{c1}$ and $\chi_{c2}$
  asymmetries compared to the 
projected statistical errors will allow 
 to  draw 
conclusions on the polarized gluon distribution function.

In addition, the measurement of $\chi_{c1}$ and $\chi_{c2}$ asymmetries  
provides a good opportunity to discriminate between two 
possible mechanisms of heavy quarkonium production: the NRQCD factorization 
approach and the Color Evaporation Model (CEM) \cite{CEM}. 
As can be seen from Fig. 3b, the $\chi_{c2}$ production asymmetry is negative
at smaller values of transverse momentum where the 
gluon-gluon fusion subprocess
contribution to $\Delta\sigma$ is dominant. In spite of  the positive
 contribution of the color octet
state $^3S_1^{(8)}$ to  $\Delta\sigma$,  
  the $\chi_{c2}$ production asymmetry remains negative as in the CSM, i.e. 
 in the NRQCD approach the double spin asymmetries for $\chi_{c1}$ and
 $\chi_{c2}$ states have 
different sign. In contrast, in the CEM the asymmetries for all 
charmonium states 
are expected to be  the same. Such a comparison is possible even 
when shape and size of the polarized gluon distribution in the proton are not
yet known.
The only requirement is that the gluon polarization  is  large enough 
to generate 
observable $\chi_{c1}$ and $\chi_{c2}$ asymmetries.

For  comparison we have also calculated the expected double-spin
asymmetries for the   production of $\chi_{c1}$ and $\chi_{c2}$ states 
at RHIC energies.
The results are given in Fig. 5 for the c.m.s. energy $\sqrt{s}=200$ GeV.
The statistical errors are calculated with the anticipated
integrated luminosity of $320$ $pb^{-1}$ assumming $100\%$ efficiency and 
$P_B P_T\simeq0.5$. 
As  can be seen by comparing  Fig. 3 and Fig. 5,  the expected asymmetries
 decrease with
increasing  c.m.s. energy for a given  parameterization of polarized
parton distribution functions. Moreover, the anticipated integrated 
luminosities at HERA-$\vec N$ and RHIC lead to a smaller discrimination 
power at higher energy.
\begin{figure}[ht]
\vspace{-4mm}
\centering
\begin{minipage}[c]{7.5cm}
\centering
\epsfig{file=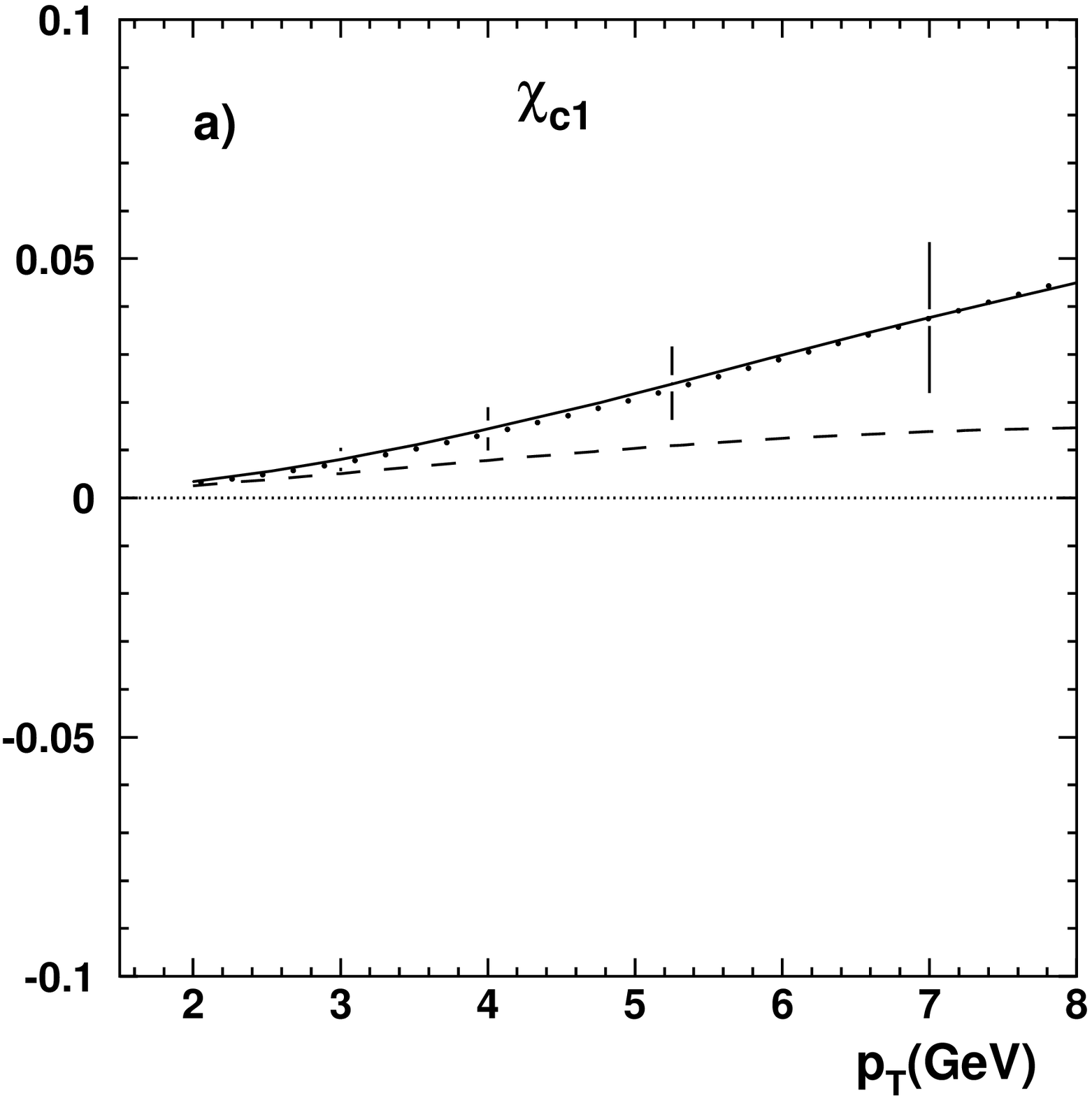,width=7.5cm}
\end{minipage}
\hspace*{0.5cm}
\begin{minipage}[c]{7.5cm}
\centering
\epsfig{file=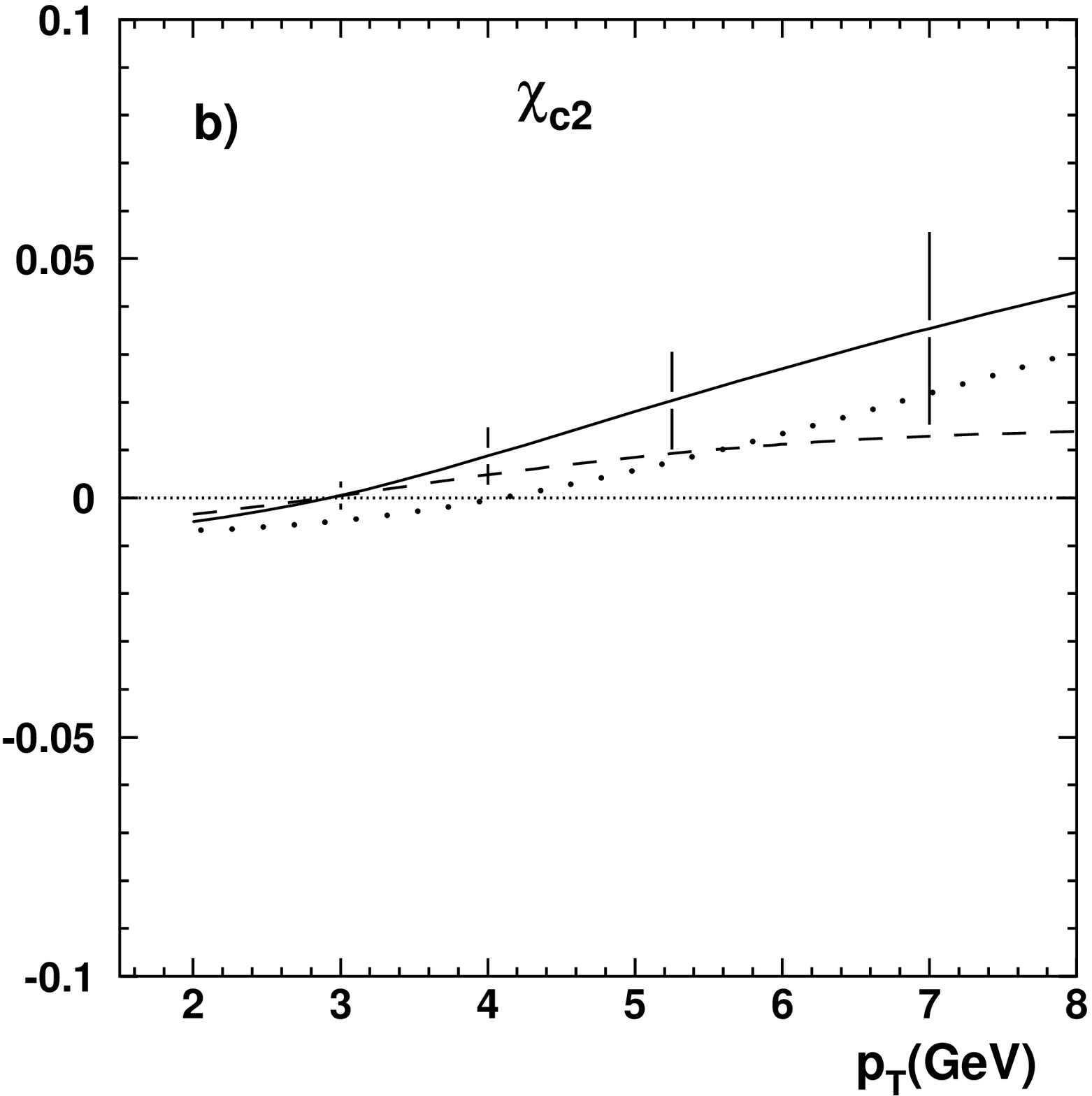,width=7.5cm}
\end{minipage}
\caption{\small
  a) $\chi_{c1}$ and  b) $\chi_{c2}$ production  asymmetries
 versus transverse momentum 
at  $\sqrt{s}=200$ GeV.
 The solid (dashed) line correspond
to set A (B) of the NLO GS parametrization.
The dotted line represents the expected asymmetry calculated only 
in CSM for the set A of the NLO parametrization.
}

\end{figure}

One of the main parameters of the factorization approach is the mass of the
charm quark. As in the case of $J/\psi$ production \cite{TT},
the expected asymmetries for $\chi_{cJ}$ states are
practically insensitive to the value of the  charm quark mass.
Therefore, the double spin asymmetry in  $\chi_{cJ}$   production,
unlike the cross section, should be free from uncertainties caused
by the unknown mass of  intermediate color octet states. Moreover, the 
asymmetries do not strongly depend on the renormalization scale,
 unlike the production cross sections \cite{BK}.
The effects of intrinsic transverse momentum smearing on
the asymmetries in the production of charmonium states were also
investigated. Such an effect
 is very important for the cross section of $J/\psi$ photoproduction 
 at HERA collider energies \cite{SMS}, however its influence 
  to asymmetries is far below the size of the projected 
statistical errors shown in Figs. 3 and 5.

\vspace{-3mm}

\section{Conclusions}

\vspace{-3mm}

Double spin asymmetries expected in hadroproduction of 
heavy quarkonium $P$-wave states  in polarized proton-proton
collisions were investigated.
To reduce the contribution from possible higher twist
 corrections \cite{VHBT} 
  the production of    $\chi_{cJ}$  mesons was considred
 at non-zero transverse momenta, $p_T>1.5$ GeV unlike the calculations of
\cite{GM,Don}, where only the lowest
order subprocesses $2\to1$ were taken into account.

 The size of the expected asymmetries in conjunction with the
projected statistical uncertainties  at HERA-$\vec N$ ($\sqrt{s}\simeq40$ GeV) 
will allow  to distinguish between different parametrizations for polarized 
parton distribution functions, specifically between  
the NLO GS sets A and  B  \cite{GSnew}. 
At RHIC, higher than hitherto planned integrated luminosities
would be required to do so.

The  asymmetries for the production of the  $\chi_{c1}$ and $\chi_{c2}$
states
 in the  transverse momentum range $1.5<p_t<3$ GeV exhibit different 
signs in the NRQCD factorization approach.
In contrast to this, the color evaporation model predicts identical 
 double 
spin asymmetries  for all charmonium states.
This  should allow to discriminate between these  different mechanisms
 of heavy 
quarkonia production if the gluon polarization is sufficiently large  and
 $\chi_c$  asymmetries become observable.

We are grateful to Oleg Teryaev for helpful discussions. We thank
J.~Bl\"umlein for careful reading of the manuscript.
A.T. acknowledges  the  support  by the Alexander von
Humboldt Foundation.

\end{document}